\documentclass[aps,prb,reprint,superscriptaddress,showpacs]{revtex4-1}

\usepackage{graphicx,psfrag}
\usepackage[latin1]{inputenc}
\usepackage{natbib}
\usepackage{amsmath, amsthm, amssymb}
\usepackage{color}
\usepackage{comment}
\usepackage{setspace}
\usepackage{gensymb}
\usepackage{listings}
\usepackage{hyperref}

\newcommand{\mb}{\mathbf}
\newcommand{\notes}[1]{}

\newcommand{\beq}{\begin{equation}}
\newcommand{\eeq}{\end{equation}}
\newcommand{\beqnn}{\begin{equation*}}
\newcommand{\eeqnn}{\end{equation*}}
\newcommand{\beqas}{\begin{eqnarray*}}
\newcommand{\eeqas}{\end{eqnarray*}}
\newcommand{\beqa}{\begin{eqnarray}}
\newcommand{\eeqa}{\end{eqnarray}}

\begin{document}

\title{Observation of droplet soliton drift resonances in a spin-transfer-torque nanocontact to a ferromagnetic thin film}
\date{\today}
\author{S. Lend\'{i}nez}
\email{sergi.lendi@ubxlab.com}
\affiliation{Grup de Magnetisme, Departament de F\'{i}sica Fonamental, Universitat de Barcelona, Spain}
\author{N. Statuto}
\affiliation{Grup de Magnetisme, Departament de F\'{i}sica Fonamental, Universitat de Barcelona, Spain}
\author{D. Backes}
\author{A.D. Kent}
\affiliation{Department of Physics, New York University, New York, New York 10003, USA}
\author{ F. Macià}
\affiliation{Grup de Magnetisme, Departament de F\'{i}sica Fonamental, Universitat de Barcelona, Spain}


\begin{abstract}

Magnetic droplet solitons are non-linear dynamical modes that can be excited in a thin film with perpendicular magnetic anisotropy with a spin-transfer-torque. Although droplet solitons have been proved to be stable with a hysteretic response to applied currents and magnetic fields at low temperature, measurements at room temperature indicate less stability and reduced hysteresis width. Here, we report evidence of droplet soliton drift instabilities, leading to drift resonances, at room temperature that explains their lower stability. Micromagnetic simulations show that the drift instability is produced by an effective field asymmetry in the nanocontact region that can have different origins.

\end{abstract}

\maketitle

\label{intro}

Nanometer scale point contacts to ferromagnetic thin films with a free magnetic layer (FL) and a fixed spin-polarizing magnetic layer (PL) \cite{Tsoi,Kiselev,Rippard2004,Bonettiprl2010} are known as spin torque nano-oscillators (STNO). In these nanocontacts, it is possible to excite steady state spin-precession by compensating the dissipation due to magnetic damping with spin-transfer torques from an electrical current. A large enough current density provides sufficient spin-transfer-torque effect to induce magnetic excitations \cite{Slonczewski1996,berger1996}. Such excitations are the building blocks for new applications \cite{stefanoBook, julie2014}
beyond binary computing and memory devices \cite{iopMacia}.

\begin{figure}[htbp!]
\includegraphics[width=\columnwidth]{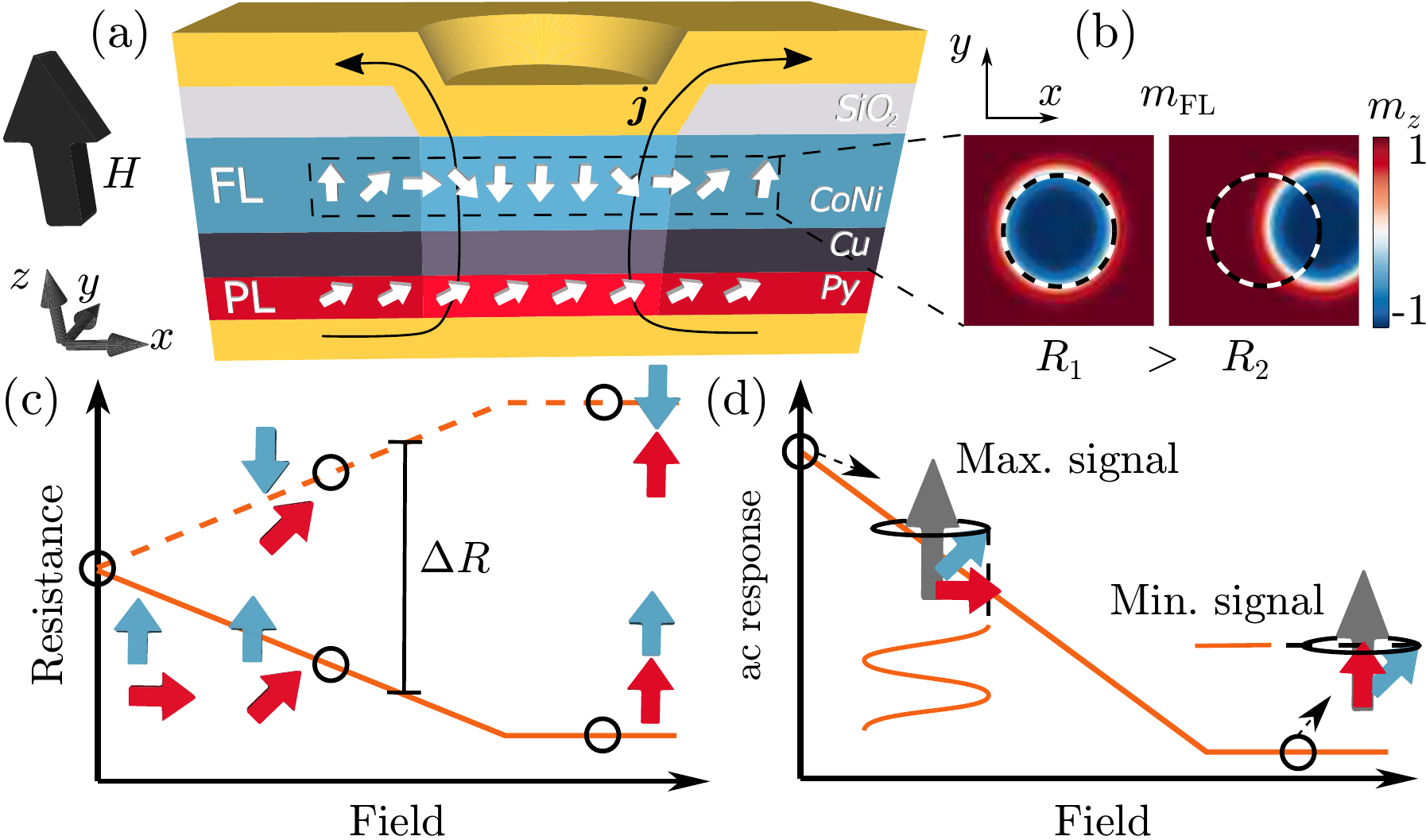}
\caption{(Color Online) (a) Schematic of the STNO. An electrical current flows through a nanocontact (80-150 nm in diameter) to a thin ferromagnetic layer (free layer, FL) and a spin-polarizing layer (PL). (b) Schematic configuration of a droplet soliton centered beneath the nanocontact (left image) and the same soliton moved sideways (right image). (c) Schematic dc resistance and (d) ac response, showing the spin configuration of PL and FL for a  nanocontact as a function of the out-of-plane field.}
\label{fig1}
\end{figure}

In layers with perpendicular magnetic anisotropy (PMA) the spin-transfer torques are predicted to lead to \emph{dissipative droplet solitons} (herafter simply refered as droplet solitons) \cite{Hoefer2010}. The spin torque compensates the damping of the material, hence directly relating these kind of solitons to the conservative magnon droplets in uniaxial (easy-axis type) ferromagnets with no damping \cite{Ivanov1977,Kosevich1990}. Droplet solitons are dynamical excitations that localize in the contact region \cite{Hoefer2010,backes2015} and are predicted to have their magnetization almost completely reversed relative to the film magnetization outside the contact [see Fig.\ \ref{fig1}(a)]. Recent experiments have shown that at low temperature the predicted reversal of the magnetization occurs \cite{DropletMacia2014}. However, room temperature measurements proved there is an abrupt threshold in both current and field at which excitations occur and showed that spin precession frequencies were below the FMR frequency \cite{scienceDroplet2013,physicaB2013, akermanJAP2014}, but there was no evidence for fully reversed magnetization beneath the nanocontact.

Here, we report observation of drift instabilities \cite{Hoefer2010, Hoefer2012} in droplet-soliton excitations at room temperature. We studied the effect of magnetic fields and current densities on the droplet dynamics. Our modeling shows that drift instabilities---caused by small asymmetries in the system, namely a variation of either the effective field or the magnetic anisotropy in the nanocontact region---create a low-frequency dynamics (hundreds of MHz) in the soliton that can be detected electrically. Further, through simulations we have confirmed that drift instabilities force the soliton to shift out of the nanocontact region and to annihilate while at the same time a new soliton emerges at the center of the contact resulting in a drift resonance.

Our samples were fabricated with a free layer (FL) with PMA and an in-plane fixed polarizing layer (PL). The FL is a 4 nm thick multilayer of CoNi, and the PL layer, a 10 nm thick Ni$_{80}$Fe$_{20}$ Permalloy (Py). A spacing layer of copper (Cu) magnetically decouples the two layers. SiO$_2$ was used as a dielectric and Au for the contacts and pads. Circular contacts of different diameters were patterned through the dielectric by electron-beam lithography with diameters ranging from 80 to 150 nm. The effective anisotropy field from the CoNi multilayer was measured through ferromagnetic resonance spectroscopy (FMR) and resulted in $\mu_0 H_a(=\mu_0 (H_K-M_S))=0.25$~T, with $\mu_0$ the permeability of free space \cite{Macia:jmmm2012}.

The electrical response in our measurements is associated with the giant magnetoresistance effect; the in-plane polarizer allows us to detect variations of the free layer magnetization in the contact region. The resistance of the nanocontacts depends on the relative magnetization orientation between the FL $\mb{m}_{\text{FL}}$, and PL $\mb{m}_{\text{PL}}$, being the fractional change $\mathrm{MR}=\overline{R}_0(1-\mb{m}_{\text{FL}} \cdot \mb{m}_{\text{PL}})/2$, with $\overline{R}_0=(R_\mathrm{AP}-R_\mathrm{P})/R_P$, $R_\mathrm{AP,P}$ being the resistance between the device antiparallel (AP) and parallel (P) magnetization states. An applied magnetic field perpendicular to the film plane tilts the Py magnetic moments out of the film plane, $\mb{m}_{\text{PL}}\cdot\mb{z}=H/M_S$ for $H<M_S$ with $\mu_0 M_S\approx 1$ T; the CoNi magnetization remains perpendicular to the film plane. Therefore, at large applied fields, $H> M_S$, the magnetization of the two layers forms a P state.

Figure \ref{fig1}(c) shows schematically the point contact resistance as a function of the applied out-of-plane field; the resistance decreases linearly with the applied field for $H<M_S$ and saturates when $H\geq M_S$. The measured maximum overall MR of the STNO ($\overline{R}_0=0.2$ \%) corresponds to twice the difference between resistance at zero field, where the magnetization of the PL and FL are orthogonal, and the resistance at a large field ($H>M_S$), where the PL and FL are in a P state, divided by the resistance of the P state. The dashed red curve in Fig.~\ref{fig1}(c) illustrates the expected resistance for a reversed FL magnetization (i.e., magnetization opposite to the applied field). This curve is obtained by reflecting the measured resistance about the horizontal line, $R(H=0)$.

In our experiments we detected dc variations of the resistance that marked the onset of droplet soliton excitations, the FL magnetization orientation changes at a threshold current while the PL magnetization remains fixed. We were also able to detect oscillations in the resistance, and thus in the voltage response, associated to the oscillations of the in-plane component of the FL magnetization. We note that the latter produces a voltage signal for fields lower than the saturation field of the PL (i.e., when the polarizing layer has a component of magnetization in the film plane) and vanishes when the PL is saturated in the same direction as the FL magnetization [see Fig.\ \ref{fig1}(d)]. Finally, we also measured a low-frequency signal with a characteristic timescale of hundreds of MHz (similar to domain wall resonances \cite{Argyle1984,Chikazumi,Saitoh2004}). The low-frequency signal does not vanish when the PL is saturated, contrary to the high-frequency precession of the FL magnetization [see Fig.\ \ref{fig1}(d)], suggesting a drift resonance of the overall soliton structure beneath the nanocontact area.

We first studied the onset and annihilation of droplet excitations through dc-measurements of resistance. We fixed an applied field out-of-the film plane and swept the applied current. These measurements show an abrupt increase in the resistance when the droplet forms when sweeping the current up---and an abrupt decrease as the droplet annihilates when sweeping the current down.

\begin{figure}[ht]
\includegraphics[width=\columnwidth]{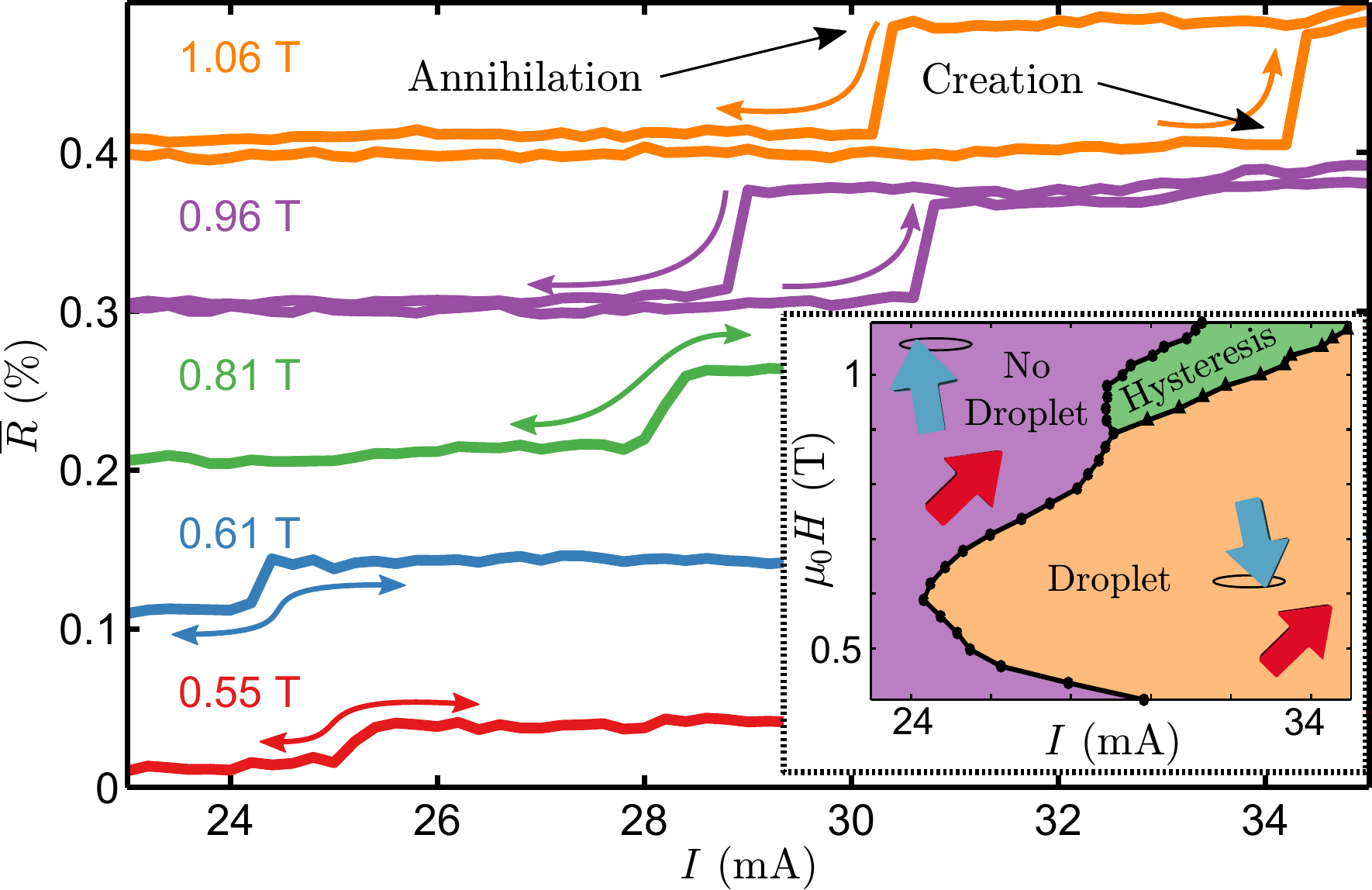}
\caption{(Color Online) Measured normalized resistance $\overline{R}=R/R_\mathrm{P}$ as a function of applied current for fields ranging from 0.5 to 1.1 T. (Inset) Stability map of the droplet soliton: on the hysteretic area, triangles show creation of the droplet, and dots annihilation.}
\label{fig2}
\end{figure}

Figure \ref{fig2} shows the resistance curves as a function of the current at different fields. At each field, the current was swept up to 35 mA and then back down to 0. Although we detected the onset of the excitations at fields $0.5\,\mathrm{T}< \mu_0 H < 0.9\, \mathrm{T}$, it is not until larger field values, $\mu_0 H > 0.9\, \mathrm{T}$, that we observed hysteresis phenomena: while sweeping the current up, the droplet creation occurs at higher currents than the annihilation when sweeping it down. In the inset of Fig.\ \ref{fig2} we plot a state map representing creation and annihilation currents for all measured fields.

The hysteresis was already observed and characterized at low temperature \cite{DropletMacia2014}. However, room temperature measurements showed much smaller hysteretic effects \cite{scienceDroplet2013}. We note here that we measured several devices and although the onset maps for the droplet excitations were almost identical, the hysteretic responses were considerably different: all samples showed hysteresis at larger fields, $\mu_0 H > 0.9\, \rm{T}$, but the size of the hysteresis varied between 0.5 and 5 mA.

The size of the jumps in resistance---representing the difference between no excitation and droplet excitation, $\Delta R$---are field dependent, as the angle of the PL magnetization increases with field. The measured maximum change in resistance is $\Delta \overline{R}\equiv\Delta R/R_\mathrm{P}\approx 0.08$ \% for fields above the saturation of the FL magnetization ($\approx 1$ T), smaller than the maximum total change of $\overline{R}_0=0.2$ \% for the AP configuration (being only $\Delta \overline{R}/\overline{R}_0\approx 1/3$). Thus we conclude that the spins are not fully reversed during the measurement, which is a time-average measurement of the contact resistance. One hypothesis is that the magnetization of the FL precesses at an effective angle of about 70 degrees in the contact region; another possibility is that the excitation is smaller than the nanocontact size, or that the excitation moves beneath the contact during the measurement time (drift instabilities \cite{Hoefer2010,Hoefer2012}).

\begin{figure}[ht]
\includegraphics[width=\columnwidth]{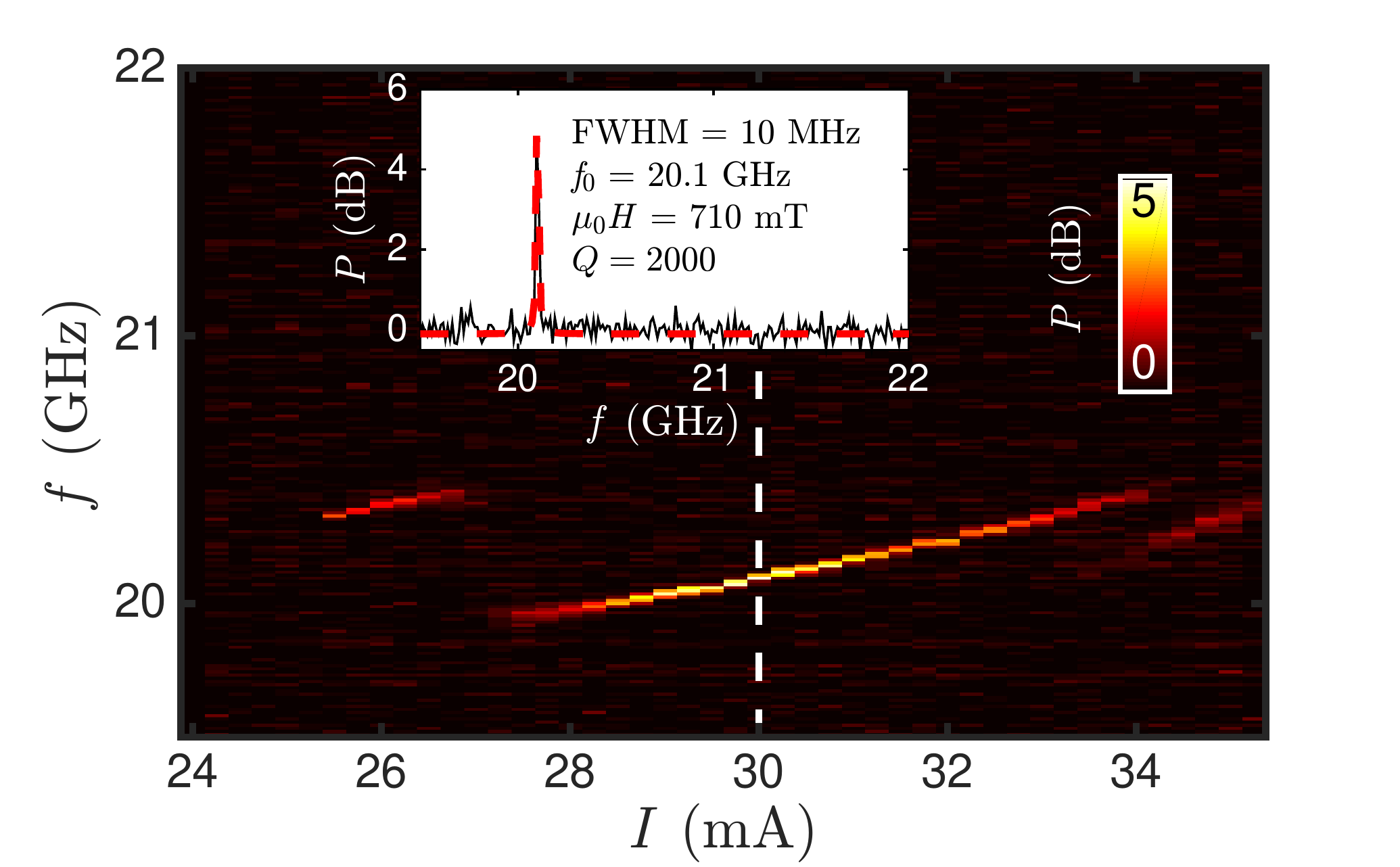}
\caption{(Color Online) High-frequency spectra as a function of applied current for a field of 710 mT. (Inset) Signal, in dB above noise, at a fixed current of $I=30\,\rm{mA}$ (white dashed line). The fitted data (red dashed line) shows a narrow peak with a FWHM = 10 MHz, and quality factor $Q\approx2000$.}
\label{fig3}
\end{figure}

We have measured the high-frequency resistance signal of the droplet excitations. Figure \ref{fig3} shows high-frequency spectra at a field of $\mu_0 H=710$ mT. At the onset current the characteristic frequency of the droplet excitation is $f \approx$ 20.3 GHz, below the corresponding FMR frequency (27.5 GHz, measured in the same film), and close to the Zeeman frequency ($\gamma\mu_0H=19.9$~GHz, $\gamma$ being the gyromagnetic ratio). As the current increases, the frequency  increases (there is a blueshift) until it jumps to a lower frequency where it continues shifting with the same trend. In the inset of Fig.\ \ref{fig3} we have plotted the power spectra at a fixed applied current of $I=30$ mA to show that the quality factor of the droplet oscillations is good ($Q\approx2000$), having peak widths of about 10 MHz. Resonance frequencies increase with the applied field having a nominal value always below the FMR (between 5 and 8 GHz). We measured similar current-dependence spectra at different applied fields. We note here that with increasing the applied field the microwave signal vanishes because the magnetization of the PL saturates perpendicular to the film plane in the same direction as the FL [see Fig.\ \ref{fig1}(d)].

\begin{figure}[htbp!]
\includegraphics[width=\columnwidth]{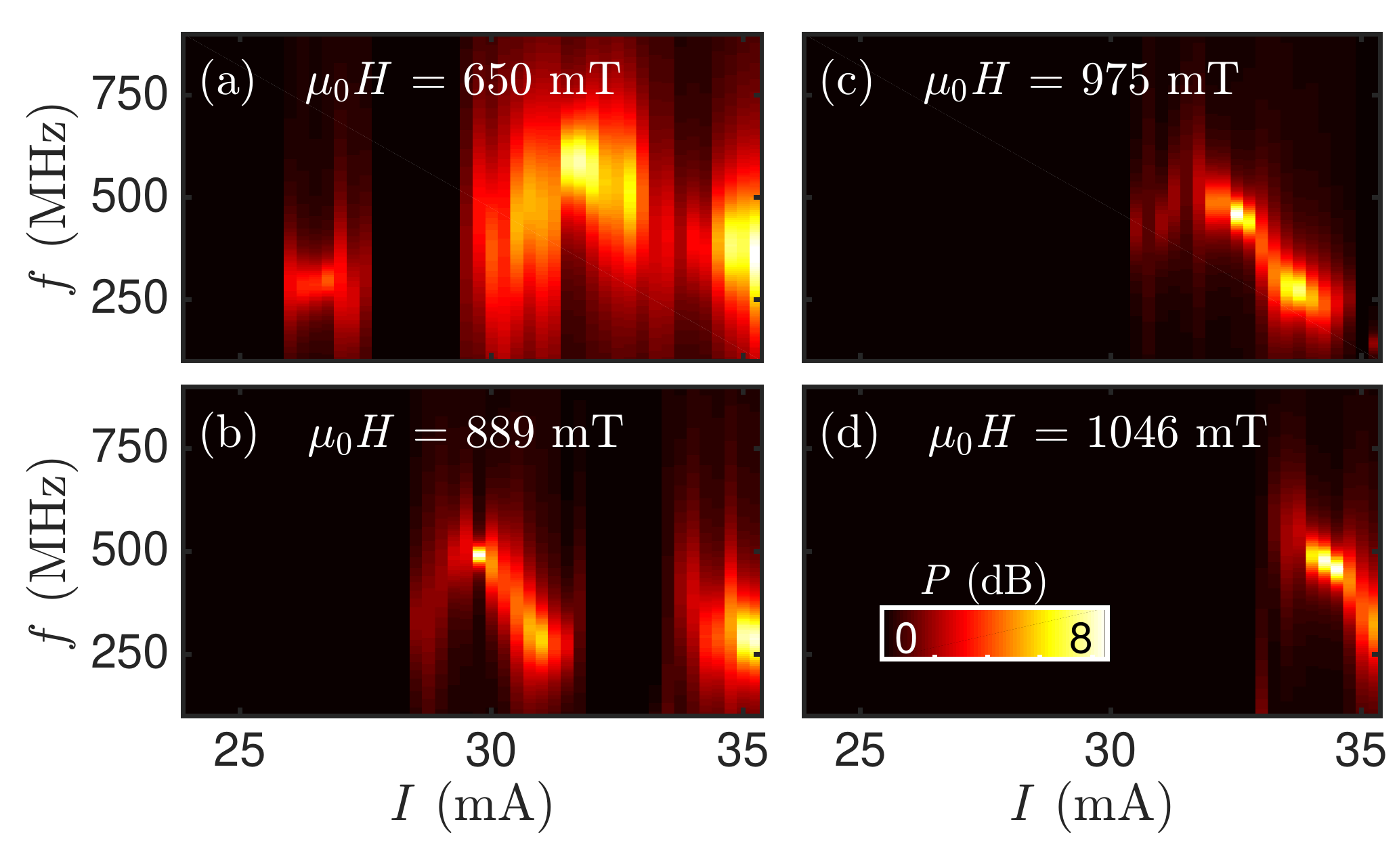}
\caption{(Color Online) Low frequency spectra as a function of applied current for fields of (a) 650 mT, (b) 889 mT, (c) 975 mT, and (d) 1046 mT.}
\label{fig4}
\end{figure}

Next, we measured voltage signal at much lower frequencies (hundreds of MHz) in order to find oscillatory dynamics related to the droplet soliton as a whole object. Along with the creation of droplet excitations, we found a strong and broad oscillating signal at about 300 MHz with a weak dependence on applied field and current. The low-frequency signal appears together with the step in resistance and we associate it with the creation of the droplet excitation, see Fig. \ref{fig4}. All samples in the droplet state showed this low frequency voltage signal in the range of 100-800 MHz but the shape of the peaks were considerably different, having a well defined peak structure in some cases and a much broader structure in others. Further, we measured hysteresis in the appearance of the low frequency signal along with the hysteresis in resistance. At applied fields that do not saturate the PL we observe different modes with increasing the current [see Fig.\ \ref{fig4}(a),(b)] and, interestingly, at fields that saturate the PL [see Fig.\ \ref{fig4}(c),(d)] the signal remains strong having a single peak. The magnetoresistance signal caused by the FL spin precession around the effective field vanishes when the PL is saturated; we thus infer that the origin of the low-frequency signal is not a precessing dynamics of the FL magnetization but motion of the whole droplet.

Moderate in-plane fields increase the power of the low-frequency signal. We measured both dc and low-frequency MR signals as a function of the angle of the applied field, 0 degrees being a field perpendicular to the film, and obtained that the droplet annihilates and does not form again at angles above $\pm15$ degrees, depending on the applied current. The dc resistance signal stayed at a higher value (corresponding to droplet state) constant as a function of the angle until it drops to the lower value (no droplet) at a certain angle; the low-frequency signal increased with the angle (increased with the in plane field). (See Supplemental Material S1\footnote{See Supplemental Material for additional measurements on other samples and simulations}).

\label{micro}

We modeled the nanocontact and the droplet excitations with micromagnetic simulations using the open-source MuMax code \cite{mumax3}, and performed parallel calculations with a graphics card with 2048 processing cores.

The material parameters taken were those we determined experimentally with FMR (see above). A circular nanocontact of 150 nm in diameter was modeled to fit the nominal diameter of the measured samples. We used a damping parameter $\alpha=0.03$ and adjusted the spin torque efficiency to obtain a droplet onset map similar to the measured in Fig.\ \ref{fig2}. We also considered the effects of the Oersted fields. (Full codes are available in the Supplemental Material S2\cite{Note1}.) Additionally, we simulated time frames of hundreds of nanoseconds in order to resolve the slow motion of the solitons.

\begin{figure}[htbp!]
\includegraphics[width=\columnwidth]{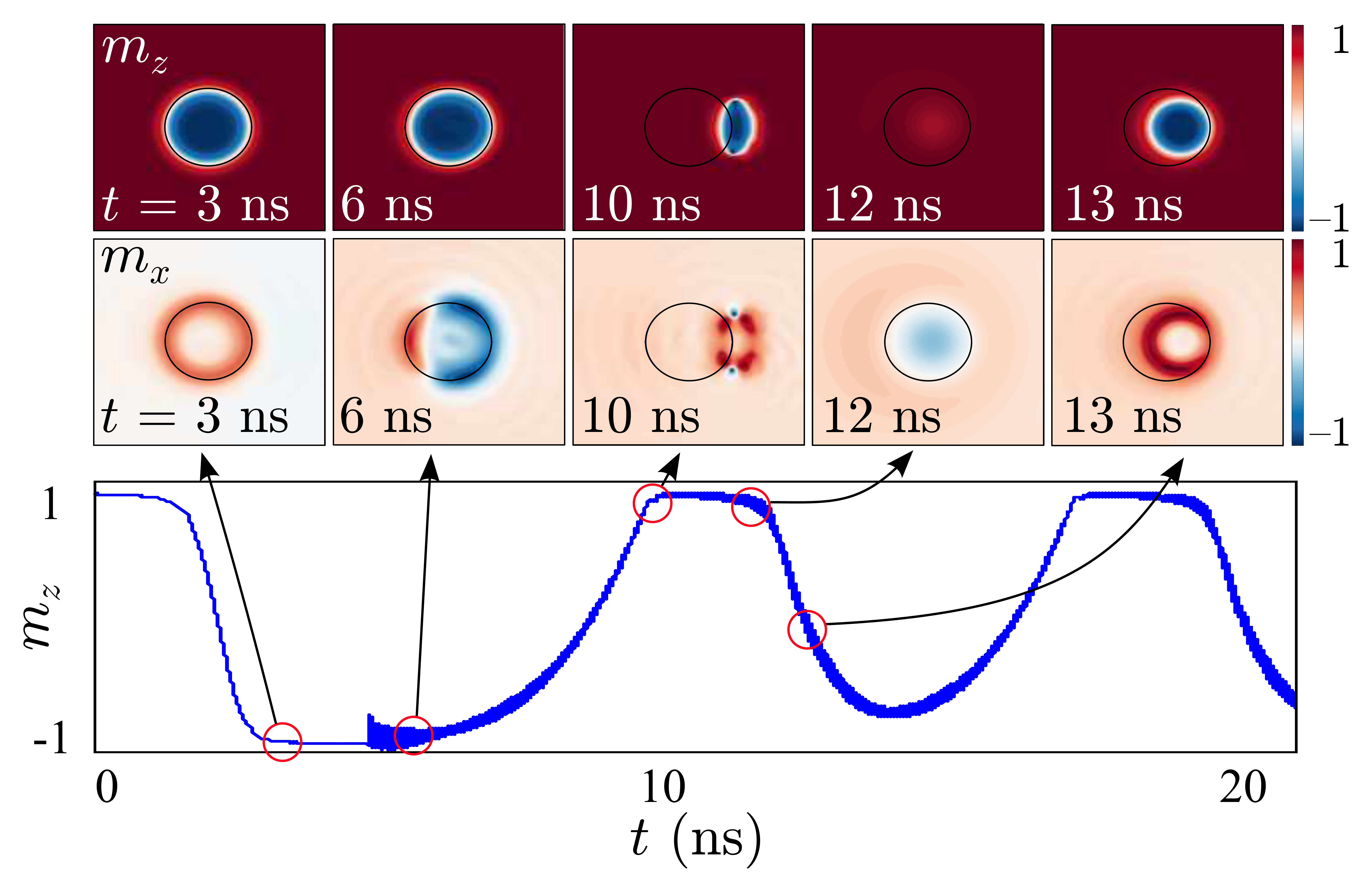}
\caption{(Color Online) Time evolution of the droplet soliton in an applied field of 1.1 T perpendicular to the film plane first and with an additional in-plane field ($y$ direction) of 0.15 T for times $t\geq5$ ns. The upper panels show magnetization maps for $m_z$, and $m_x$, at particular times of the simulation. Images correspond to a 400 $\times$ 400 nm$^2$ field view. The contact region is outlined in black. The lower panel shows the time evolution of the perpendicular component of the magnetization $m_z$, averaged in the nanocontact area.}
\label{sim}
\end{figure}

Our simulations show that droplet solitons form and annihilate at different critical current values. The hysteresis describes the stability of the object (with no temperature effects) and corresponds to a larger value than twice the anisotropy ($2\,\mu_0H_a=0.5$ T) \cite{DropletMacia2014}. Our experiments showed however a much smaller hysteresis, $\approx 0.1\ \mathrm{T}$.

Under these conditions, our simulations did not show drift instabilities and thus no low-frequency dynamics. However, we found that a small in-plane field ($\mu_0 H\sim 0.1$ T) causes a dramatic change: droplet excitations shift in the direction perpendicular to the applied in-plane field and annihilate as they get out of the nanocontact region because of the damping. Immediately, a new droplet soliton is created beneath the nanocontact \cite{Hoefer2010}. This process has a frequency, for the simulated parameters, of about 150 MHz and hardly depends on the out of plane field and applied current.

Figure \ref{sim} shows the evolution of the droplet soliton in an applied field of 1.1 T perpendicular to the film plane and then with an additional in plane field of 0.15 T (equivalent to a magnetic field with an angle of $\theta\approx$~8~degrees). The upper panels show magnetization maps, for the components perpendicular to the plane, $m_z$, and in the plane, $m_x$, at particular times of the simulation. The lower panel shows the time evolution of $m_z$ in the nanocontact area. During the first 5 ns we apply only a perpendicular magnetic field of 1.1 T and observe how a droplet forms having all spins precessing in phase (see panels for $t=3$~ns). At $t>5$~ns we apply an in-plane field, 0.15 T, in the $y$ direction. This creates a drift instability, an imbalance in the precession phases that shifts the droplet in the $x$ direction (perpendicular to the applied field) until it annihilates (see panels for $t=$ 6 and 10~ns). At 12~ns it appears that the droplet has dissipated but a new excitation is being created (see panels for $t=$ 12 and 13~ns). We note that the time average of $m_z$ beneath the nanocontact---that is the measurable quantity using any dc technique---is only a 36~\% of the total, or equivalent to a precession angle of 73 degrees \cite{backes2015}.

In order to understand why we have observed low-frequency dynamics even when the applied field was perpendicular to the film we introduced asymmetric parameters in the simulations (see Supplemental Material S2 \cite{Note1}). We found that a variation in the anisotropy of only $1\ \%$ between the two halves of a nanocontact produces the same effect---with almost the same annihilation and creation frequency. In general we found that any asymmetry in the effective field causes a drift instability and results in an oscillatory signal (drift resonance) of hundreds of MHz corresponding to annihilation and creation of the soliton excitation.

Although it seems counter intuitive that hysteresis can exist when the droplet soliton is being created and annihilated, hysteresis still appears. However, the size in field of the hysteresis cycle reduces considerably in our micromagnetic simulations (from $\sim$0.6 T to $\sim$0.1~T). The reason the soliton presents certain stability even in the presence of drift instabilities is that when a soliton moves away from the contact area, the magnetization beneath the nanocontact is still precessing at a some finite angle.

\label{disc}

Drift instabilities were described as a consequence of magnetostatic interactions between the effective dipole moment of the droplet and an effective field gradient, associated with the Oersted field \cite{Hoefer2010} or with differences in the anisotropy. A small in-plane field or a small gradient in the effective field creates an asymmetric landscape that dephases the precession of magnetization between edges in the soliton boundaries resulting in a magnetic force that acts on the soliton and shifts it. Some experiments have shown the presence of side bands in the precession frequency of the soliton \cite{scienceDroplet2013} suggesting that the soliton might be undergoing drift instabilities. Micromagnetic studies have also predicted the side bands and attributed them to drift instabilities \cite{scienceDroplet2013,puliafito2014}.

Our direct observation of the low frequency dynamics proves the existence of drift instabilities and explains why dc measurements fail to measure full magnetization reversal---showing on average just a fraction of the magnetization reversed.

We note that the measured high-frequency dynamics, associated with the spin precession, has a blueshift with the applied current, different to predictions and micromagnetic results. We cannot explain it with our model and we attribute such effect to the appearance of an effective field perpendicular to the film plane due to the applied current (likely an Oersted field effect from the leads). The overall change is always smaller than 500~MHz that would correspond to a magnetic field of 20~mT.

In conclusion, we have observed and measured drift resonances in magnetic droplet solitons and have proved that droplet solitons exist and are stable at room temperature. We suggest that the drift instability is produced by an effective field asymmetry in the nanocontact region that can have different origins.

\begin{acknowledgments}
F.M. acknowledges support from Catalan Government through COFUND-FP7. Research at the UB was partially supported by Spanish Government Project No. MAT2011-23698. Research at NYU was supported by NSF-DMR-1309202.
\end{acknowledgments}

\bibliography{Droplet_arx}

\begin{thebibliography}{25}%
\makeatletter
\providecommand \@ifxundefined [1]{%
 \@ifx{#1\undefined}
}%
\providecommand \@ifnum [1]{%
 \ifnum #1\expandafter \@firstoftwo
 \else \expandafter \@secondoftwo
 \fi
}%
\providecommand \@ifx [1]{%
 \ifx #1\expandafter \@firstoftwo
 \else \expandafter \@secondoftwo
 \fi
}%
\providecommand \natexlab [1]{#1}%
\providecommand \enquote  [1]{``#1''}%
\providecommand \bibnamefont  [1]{#1}%
\providecommand \bibfnamefont [1]{#1}%
\providecommand \citenamefont [1]{#1}%
\providecommand \href@noop [0]{\@secondoftwo}%
\providecommand \href [0]{\begingroup \@sanitize@url \@href}%
\providecommand \@href[1]{\@@startlink{#1}\@@href}%
\providecommand \@@href[1]{\endgroup#1\@@endlink}%
\providecommand \@sanitize@url [0]{\catcode `\\12\catcode `\$12\catcode
  `\&12\catcode `\#12\catcode `\^12\catcode `\_12\catcode `\%12\relax}%
\providecommand \@@startlink[1]{}%
\providecommand \@@endlink[0]{}%
\providecommand \url  [0]{\begingroup\@sanitize@url \@url }%
\providecommand \@url [1]{\endgroup\@href {#1}{\urlprefix }}%
\providecommand \urlprefix  [0]{URL }%
\providecommand \Eprint [0]{\href }%
\providecommand \doibase [0]{http://dx.doi.org/}%
\providecommand \selectlanguage [0]{\@gobble}%
\providecommand \bibinfo  [0]{\@secondoftwo}%
\providecommand \bibfield  [0]{\@secondoftwo}%
\providecommand \translation [1]{[#1]}%
\providecommand \BibitemOpen [0]{}%
\providecommand \bibitemStop [0]{}%
\providecommand \bibitemNoStop [0]{.\EOS\space}%
\providecommand \EOS [0]{\spacefactor3000\relax}%
\providecommand \BibitemShut  [1]{\csname bibitem#1\endcsname}%
\let\auto@bib@innerbib\@empty
\bibitem [{\citenamefont {Tsoi}\ \emph {et~al.}(2000)\citenamefont {Tsoi},
  \citenamefont {Jansen}, \citenamefont {Bass}, \citenamefont {Chiang},
  \citenamefont {V.},\ and\ \citenamefont {Wyder}}]{Tsoi}%
  \BibitemOpen
  \bibfield  {author} {\bibinfo {author} {\bibfnamefont {M.}~\bibnamefont
  {Tsoi}}, \bibinfo {author} {\bibfnamefont {A.~G.~M.}\ \bibnamefont {Jansen}},
  \bibinfo {author} {\bibfnamefont {J.}~\bibnamefont {Bass}}, \bibinfo {author}
  {\bibfnamefont {W.-C.}\ \bibnamefont {Chiang}}, \bibinfo {author}
  {\bibfnamefont {T.}~\bibnamefont {V.}}, \ and\ \bibinfo {author}
  {\bibfnamefont {P.}~\bibnamefont {Wyder}},\ }\href@noop {} {\bibfield
  {journal} {\bibinfo  {journal} {Nature}\ }\textbf {\bibinfo {volume} {406}},\
  \bibinfo {pages} {46} (\bibinfo {year} {2000})}\BibitemShut {NoStop}%
\bibitem [{\citenamefont {Kiselev}\ \emph {et~al.}(2003)\citenamefont
  {Kiselev}, \citenamefont {Sankey}, \citenamefont {Krivorotov}, \citenamefont
  {Emley}, \citenamefont {Schoelkopf}, \citenamefont {Buhrman},\ and\
  \citenamefont {Ralph}}]{Kiselev}%
  \BibitemOpen
  \bibfield  {author} {\bibinfo {author} {\bibfnamefont {S.~I.}\ \bibnamefont
  {Kiselev}}, \bibinfo {author} {\bibfnamefont {J.~C.}\ \bibnamefont {Sankey}},
  \bibinfo {author} {\bibfnamefont {I.~N.}\ \bibnamefont {Krivorotov}},
  \bibinfo {author} {\bibfnamefont {N.~C.}\ \bibnamefont {Emley}}, \bibinfo
  {author} {\bibfnamefont {R.~J.}\ \bibnamefont {Schoelkopf}}, \bibinfo
  {author} {\bibfnamefont {R.~A.}\ \bibnamefont {Buhrman}}, \ and\ \bibinfo
  {author} {\bibfnamefont {D.~C.}\ \bibnamefont {Ralph}},\ }\href@noop {}
  {\bibfield  {journal} {\bibinfo  {journal} {Nature}\ }\textbf {\bibinfo
  {volume} {425}},\ \bibinfo {pages} {380} (\bibinfo {year}
  {2003})}\BibitemShut {NoStop}%
\bibitem [{\citenamefont {Rippard}\ \emph {et~al.}(2004)\citenamefont
  {Rippard}, \citenamefont {Pufall}, \citenamefont {Kaka},\ and\ \citenamefont
  {Silva}}]{Rippard2004}%
  \BibitemOpen
  \bibfield  {author} {\bibinfo {author} {\bibfnamefont {W.}~\bibnamefont
  {Rippard}}, \bibinfo {author} {\bibfnamefont {M.}~\bibnamefont {Pufall}},
  \bibinfo {author} {\bibfnamefont {S.}~\bibnamefont {Kaka}, \bibfnamefont
  {S.~Russek}}, \ and\ \bibinfo {author} {\bibfnamefont {T.}~\bibnamefont
  {Silva}},\ }\href@noop {} {\bibfield  {journal} {\bibinfo  {journal} {Phys.
  Rev. Lett}\ }\textbf {\bibinfo {volume} {92}},\ \bibinfo {pages} {027201}
  (\bibinfo {year} {2004})}\BibitemShut {NoStop}%
\bibitem [{\citenamefont {Bonetti}\ \emph {et~al.}(2010)\citenamefont
  {Bonetti}, \citenamefont {Tiberkevich}, \citenamefont {Consolo},
  \citenamefont {Finocchio}, \citenamefont {Muduli}, \citenamefont {Mancoff},
  \citenamefont {Slavin},\ and\ \citenamefont {\AA{}kerman}}]{Bonettiprl2010}%
  \BibitemOpen
  \bibfield  {author} {\bibinfo {author} {\bibfnamefont {S.}~\bibnamefont
  {Bonetti}}, \bibinfo {author} {\bibfnamefont {V.}~\bibnamefont
  {Tiberkevich}}, \bibinfo {author} {\bibfnamefont {G.}~\bibnamefont
  {Consolo}}, \bibinfo {author} {\bibfnamefont {G.}~\bibnamefont {Finocchio}},
  \bibinfo {author} {\bibfnamefont {P.}~\bibnamefont {Muduli}}, \bibinfo
  {author} {\bibfnamefont {F.}~\bibnamefont {Mancoff}}, \bibinfo {author}
  {\bibfnamefont {A.}~\bibnamefont {Slavin}}, \ and\ \bibinfo {author}
  {\bibfnamefont {J.}~\bibnamefont {\AA{}kerman}},\ }\href {\doibase
  10.1103/PhysRevLett.105.217204} {\bibfield  {journal} {\bibinfo  {journal}
  {Phys. Rev. Lett.}\ }\textbf {\bibinfo {volume} {105}},\ \bibinfo {pages}
  {217204} (\bibinfo {year} {2010})}\BibitemShut {NoStop}%
\bibitem [{\citenamefont {Slonczewski}(1996)}]{Slonczewski1996}%
  \BibitemOpen
  \bibfield  {author} {\bibinfo {author} {\bibfnamefont {J.~C.}\ \bibnamefont
  {Slonczewski}},\ }\href@noop {} {\bibfield  {journal} {\bibinfo  {journal}
  {Journal of Magnetism and Magnetic Materials}\ }\textbf {\bibinfo {volume}
  {1-2}},\ \bibinfo {pages} {L1 } (\bibinfo {year} {1996})}\BibitemShut
  {NoStop}%
\bibitem [{\citenamefont {Berger}(1996)}]{berger1996}%
  \BibitemOpen
  \bibfield  {author} {\bibinfo {author} {\bibfnamefont {L.}~\bibnamefont
  {Berger}},\ }\href@noop {} {\bibfield  {journal} {\bibinfo  {journal} {Phys.
  Rev. B}\ }\textbf {\bibinfo {volume} {54}},\ \bibinfo {pages} {9352}
  (\bibinfo {year} {1996})}\BibitemShut {NoStop}%
\bibitem [{\citenamefont {Bonetti}\ and\ \citenamefont
  {{\AA}kerman}(2013)}]{stefanoBook}%
  \BibitemOpen
  \bibfield  {author} {\bibinfo {author} {\bibfnamefont {S.}~\bibnamefont
  {Bonetti}}\ and\ \bibinfo {author} {\bibfnamefont {J.}~\bibnamefont
  {{\AA}kerman}},\ }in\ \href@noop {} {\emph {\bibinfo {booktitle}
  {Magnonics}}}\ (\bibinfo  {publisher} {Springer Berlin Heidelberg},\ \bibinfo
  {year} {2013})\ pp.\ \bibinfo {pages} {177--187}\BibitemShut {NoStop}%
\bibitem [{\citenamefont {Locatelli}\ \emph {et~al.}(2014)\citenamefont
  {Locatelli}, \citenamefont {Cros},\ and\ \citenamefont
  {Grollier}}]{julie2014}%
  \BibitemOpen
  \bibfield  {author} {\bibinfo {author} {\bibfnamefont {N.}~\bibnamefont
  {Locatelli}}, \bibinfo {author} {\bibfnamefont {V.}~\bibnamefont {Cros}}, \
  and\ \bibinfo {author} {\bibfnamefont {J.}~\bibnamefont {Grollier}},\
  }\href@noop {} {\bibfield  {journal} {\bibinfo  {journal} {Nature materials}\
  }\textbf {\bibinfo {volume} {13}},\ \bibinfo {pages} {11} (\bibinfo {year}
  {2014})}\BibitemShut {NoStop}%
\bibitem [{\citenamefont {Maci\`a}\ \emph {et~al.}(2011)\citenamefont
  {Maci\`a}, \citenamefont {Kent},\ and\ \citenamefont
  {Hoppensteadt}}]{iopMacia}%
  \BibitemOpen
  \bibfield  {author} {\bibinfo {author} {\bibfnamefont {F.}~\bibnamefont
  {Maci\`a}}, \bibinfo {author} {\bibfnamefont {A.~D.}\ \bibnamefont {Kent}}, \
  and\ \bibinfo {author} {\bibfnamefont {F.~C.}\ \bibnamefont {Hoppensteadt}},\
  }\href@noop {} {\bibfield  {journal} {\bibinfo  {journal} {Nanotechnology}\
  }\textbf {\bibinfo {volume} {22}},\ \bibinfo {pages} {095301} (\bibinfo
  {year} {2011})}\BibitemShut {NoStop}%
\bibitem [{\citenamefont {Hoefer}\ \emph {et~al.}(2010)\citenamefont {Hoefer},
  \citenamefont {Silva},\ and\ \citenamefont {Keller}}]{Hoefer2010}%
  \BibitemOpen
  \bibfield  {author} {\bibinfo {author} {\bibfnamefont {M.~A.}\ \bibnamefont
  {Hoefer}}, \bibinfo {author} {\bibfnamefont {T.~J.}\ \bibnamefont {Silva}}, \
  and\ \bibinfo {author} {\bibfnamefont {M.~W.}\ \bibnamefont {Keller}},\
  }\href@noop {} {\bibfield  {journal} {\bibinfo  {journal} {Phys. Rev. B}\
  }\textbf {\bibinfo {volume} {82}},\ \bibinfo {pages} {054432} (\bibinfo
  {year} {2010})}\BibitemShut {NoStop}%
\bibitem [{\citenamefont {Ivanov}\ and\ \citenamefont
  {Kosevich}(1977)}]{Ivanov1977}%
  \BibitemOpen
  \bibfield  {author} {\bibinfo {author} {\bibfnamefont {A.}~\bibnamefont
  {Ivanov}}\ and\ \bibinfo {author} {\bibfnamefont {A.~M.}\ \bibnamefont
  {Kosevich}},\ }\href@noop {} {\bibfield  {journal} {\bibinfo  {journal} {Zh.
  Eksp. Teor. Fiz.}\ }\textbf {\bibinfo {volume} {72}},\ \bibinfo {pages}
  {2000} (\bibinfo {year} {1977})}\BibitemShut {NoStop}%
\bibitem [{\citenamefont {Kosevich}\ \emph {et~al.}(1990)\citenamefont
  {Kosevich}, \citenamefont {Ivanov},\ and\ \citenamefont
  {Kovalev}}]{Kosevich1990}%
  \BibitemOpen
  \bibfield  {author} {\bibinfo {author} {\bibfnamefont {A.~M.}\ \bibnamefont
  {Kosevich}}, \bibinfo {author} {\bibfnamefont {B.~A.}\ \bibnamefont
  {Ivanov}}, \ and\ \bibinfo {author} {\bibfnamefont {A.~S.}\ \bibnamefont
  {Kovalev}},\ }\href@noop {} {\bibfield  {journal} {\bibinfo  {journal} {Phys.
  Rep.}\ }\textbf {\bibinfo {volume} {194}},\ \bibinfo {pages} {117} (\bibinfo
  {year} {1990})}\BibitemShut {NoStop}%
\bibitem [{\citenamefont {Backes}\ \emph {et~al.}(2015)\citenamefont {Backes},
  \citenamefont {Maci{\`a}}, \citenamefont {Bonetti}, \citenamefont {Kukreja},
  \citenamefont {Ohldag},\ and\ \citenamefont {Kent}}]{backes2015}%
  \BibitemOpen
  \bibfield  {author} {\bibinfo {author} {\bibfnamefont {D.}~\bibnamefont
  {Backes}}, \bibinfo {author} {\bibfnamefont {F.}~\bibnamefont {Maci{\`a}}},
  \bibinfo {author} {\bibfnamefont {S.}~\bibnamefont {Bonetti}}, \bibinfo
  {author} {\bibfnamefont {R.}~\bibnamefont {Kukreja}}, \bibinfo {author}
  {\bibfnamefont {H.}~\bibnamefont {Ohldag}}, \ and\ \bibinfo {author}
  {\bibfnamefont {A.}~\bibnamefont {Kent}},\ }\href@noop {} {\bibfield
  {journal} {\bibinfo  {journal} {arXiv preprint arXiv:1504.00488}\ } (\bibinfo
  {year} {2015})}\BibitemShut {NoStop}%
\bibitem [{\citenamefont {Maci{\`a}}\ \emph {et~al.}(2014)\citenamefont
  {Maci{\`a}}, \citenamefont {Backes},\ and\ \citenamefont
  {Kent}}]{DropletMacia2014}%
  \BibitemOpen
  \bibfield  {author} {\bibinfo {author} {\bibfnamefont {F.}~\bibnamefont
  {Maci{\`a}}}, \bibinfo {author} {\bibfnamefont {D.}~\bibnamefont {Backes}}, \
  and\ \bibinfo {author} {\bibfnamefont {A.~D.}\ \bibnamefont {Kent}},\
  }\href@noop {} {\bibfield  {journal} {\bibinfo  {journal} {Nature
  nanotechnology}\ }\textbf {\bibinfo {volume} {9}},\ \bibinfo {pages} {992}
  (\bibinfo {year} {2014})}\BibitemShut {NoStop}%
\bibitem [{\citenamefont {Mohseni}\ \emph
  {et~al.}(2013{\natexlab{a}})\citenamefont {Mohseni}, \citenamefont {Sani},
  \citenamefont {Persson}, \citenamefont {Anh~Nguyen}, \citenamefont {Chung},
  \citenamefont {Pogoryelov}, \citenamefont {Muduli}, \citenamefont {Iacocca},
  \citenamefont {Eklund}, \citenamefont {Dumas}, \citenamefont {Bonetti},
  \citenamefont {Deac}, \citenamefont {Hoefer},\ and\ \citenamefont
  {Akerman}}]{scienceDroplet2013}%
  \BibitemOpen
  \bibfield  {author} {\bibinfo {author} {\bibfnamefont {S.~M.}\ \bibnamefont
  {Mohseni}}, \bibinfo {author} {\bibfnamefont {S.~R.}\ \bibnamefont {Sani}},
  \bibinfo {author} {\bibfnamefont {J.}~\bibnamefont {Persson}}, \bibinfo
  {author} {\bibfnamefont {T.~N.}\ \bibnamefont {Anh~Nguyen}}, \bibinfo
  {author} {\bibfnamefont {S.}~\bibnamefont {Chung}}, \bibinfo {author}
  {\bibfnamefont {Y.}~\bibnamefont {Pogoryelov}}, \bibinfo {author}
  {\bibfnamefont {P.~K.}\ \bibnamefont {Muduli}}, \bibinfo {author}
  {\bibfnamefont {E.}~\bibnamefont {Iacocca}}, \bibinfo {author} {\bibfnamefont
  {A.}~\bibnamefont {Eklund}}, \bibinfo {author} {\bibfnamefont {R.~K.}\
  \bibnamefont {Dumas}}, \bibinfo {author} {\bibfnamefont {S.}~\bibnamefont
  {Bonetti}}, \bibinfo {author} {\bibfnamefont {A.}~\bibnamefont {Deac}},
  \bibinfo {author} {\bibfnamefont {M.~A.}\ \bibnamefont {Hoefer}}, \ and\
  \bibinfo {author} {\bibfnamefont {J.}~\bibnamefont {Akerman}},\ }\href@noop
  {} {\bibfield  {journal} {\bibinfo  {journal} {Science}\ }\textbf {\bibinfo
  {volume} {339}},\ \bibinfo {pages} {1295} (\bibinfo {year}
  {2013}{\natexlab{a}})}\BibitemShut {NoStop}%
\bibitem [{\citenamefont {Mohseni}\ \emph
  {et~al.}(2013{\natexlab{b}})\citenamefont {Mohseni}, \citenamefont {Sani},
  \citenamefont {Dumas}, \citenamefont {Persson}, \citenamefont {Anh~Nguyen},
  \citenamefont {Chung}, \citenamefont {Pogoryelov}, \citenamefont {Muduli},
  \citenamefont {Iacocca}, \citenamefont {Eklund},\ and\ \citenamefont
  {Akerman}}]{physicaB2013}%
  \BibitemOpen
  \bibfield  {author} {\bibinfo {author} {\bibfnamefont {S.~M.}\ \bibnamefont
  {Mohseni}}, \bibinfo {author} {\bibfnamefont {S.~R.}\ \bibnamefont {Sani}},
  \bibinfo {author} {\bibfnamefont {R.~K.}\ \bibnamefont {Dumas}}, \bibinfo
  {author} {\bibfnamefont {J.}~\bibnamefont {Persson}}, \bibinfo {author}
  {\bibfnamefont {T.~N.}\ \bibnamefont {Anh~Nguyen}}, \bibinfo {author}
  {\bibfnamefont {S.}~\bibnamefont {Chung}}, \bibinfo {author} {\bibfnamefont
  {Y.}~\bibnamefont {Pogoryelov}}, \bibinfo {author} {\bibfnamefont {P.~K.}\
  \bibnamefont {Muduli}}, \bibinfo {author} {\bibfnamefont {E.}~\bibnamefont
  {Iacocca}}, \bibinfo {author} {\bibfnamefont {A.}~\bibnamefont {Eklund}}, \
  and\ \bibinfo {author} {\bibfnamefont {J.}~\bibnamefont {Akerman}},\
  }\href@noop {} {\bibfield  {journal} {\bibinfo  {journal} {Physica B:
  Condensed Matter}\ }\textbf {\bibinfo {volume} {435}},\ \bibinfo {pages} {84}
  (\bibinfo {year} {2013}{\natexlab{b}})}\BibitemShut {NoStop}%
\bibitem [{\citenamefont {Chung}\ \emph {et~al.}(2014)\citenamefont {Chung},
  \citenamefont {Mohseni}, \citenamefont {Sani}, \citenamefont {Iacocca},
  \citenamefont {Dumas}, \citenamefont {Anh~Nguyen}, \citenamefont
  {Pogoryelov}, \citenamefont {Muduli}, \citenamefont {Eklund}, \citenamefont
  {Hoefer}, ,\ and\ \citenamefont {Akerman}}]{akermanJAP2014}%
  \BibitemOpen
  \bibfield  {author} {\bibinfo {author} {\bibfnamefont {S.}~\bibnamefont
  {Chung}}, \bibinfo {author} {\bibfnamefont {S.~M.}\ \bibnamefont {Mohseni}},
  \bibinfo {author} {\bibfnamefont {S.~R.}\ \bibnamefont {Sani}}, \bibinfo
  {author} {\bibfnamefont {E.}~\bibnamefont {Iacocca}}, \bibinfo {author}
  {\bibfnamefont {R.~K.}\ \bibnamefont {Dumas}}, \bibinfo {author}
  {\bibfnamefont {T.~N.}\ \bibnamefont {Anh~Nguyen}}, \bibinfo {author}
  {\bibfnamefont {Y.}~\bibnamefont {Pogoryelov}}, \bibinfo {author}
  {\bibfnamefont {P.~K.}\ \bibnamefont {Muduli}}, \bibinfo {author}
  {\bibfnamefont {A.}~\bibnamefont {Eklund}}, \bibinfo {author} {\bibfnamefont
  {M.}~\bibnamefont {Hoefer}}, , \ and\ \bibinfo {author} {\bibfnamefont
  {J.}~\bibnamefont {Akerman}},\ }\href@noop {} {\bibfield  {journal} {\bibinfo
   {journal} {Jour. of Appl. Phys}\ }\textbf {\bibinfo {volume} {115}},\
  \bibinfo {pages} {172612} (\bibinfo {year} {2014})}\BibitemShut {NoStop}%
\bibitem [{\citenamefont {Hoefer}\ \emph {et~al.}(2012)\citenamefont {Hoefer},
  \citenamefont {Sommacal},\ and\ \citenamefont {Silva}}]{Hoefer2012}%
  \BibitemOpen
  \bibfield  {author} {\bibinfo {author} {\bibfnamefont {M.~A.}\ \bibnamefont
  {Hoefer}}, \bibinfo {author} {\bibfnamefont {M.}~\bibnamefont {Sommacal}}, \
  and\ \bibinfo {author} {\bibfnamefont {T.~J.}\ \bibnamefont {Silva}},\
  }\href@noop {} {\bibfield  {journal} {\bibinfo  {journal} {Phys. Rev. B}\
  }\textbf {\bibinfo {volume} {85}},\ \bibinfo {pages} {214433} (\bibinfo
  {year} {2012})}\BibitemShut {NoStop}%
\bibitem [{\citenamefont {Maci\`a}\ \emph {et~al.}(2012)\citenamefont
  {Maci\`a}, \citenamefont {Warnicke}, \citenamefont {Bedau}, \citenamefont
  {Im}, \citenamefont {Fischer},\ and\ \citenamefont {Kent}}]{Macia:jmmm2012}%
  \BibitemOpen
  \bibfield  {author} {\bibinfo {author} {\bibfnamefont {F.}~\bibnamefont
  {Maci\`a}}, \bibinfo {author} {\bibfnamefont {P.}~\bibnamefont {Warnicke}},
  \bibinfo {author} {\bibfnamefont {D.}~\bibnamefont {Bedau}}, \bibinfo
  {author} {\bibfnamefont {M.-Y.}\ \bibnamefont {Im}}, \bibinfo {author}
  {\bibfnamefont {D.~A.}\ \bibnamefont {Fischer}, \bibfnamefont {P.~Arena}}, \
  and\ \bibinfo {author} {\bibfnamefont {A.~D.}\ \bibnamefont {Kent}},\
  }\href@noop {} {\bibfield  {journal} {\bibinfo  {journal} {Jour. of Mag. Mag.
  Mat.}\ }\textbf {\bibinfo {volume} {324}},\ \bibinfo {pages} {3632} (\bibinfo
  {year} {2012})}\BibitemShut {NoStop}%
\bibitem [{\citenamefont {Argyle}\ \emph {et~al.}()\citenamefont {Argyle},
  \citenamefont {Terrenzio},\ and\ \citenamefont {Slonczewski}}]{Argyle1984}%
  \BibitemOpen
  \bibfield  {author} {\bibinfo {author} {\bibfnamefont {B.~E.}\ \bibnamefont
  {Argyle}}, \bibinfo {author} {\bibfnamefont {E.}~\bibnamefont {Terrenzio}}, \
  and\ \bibinfo {author} {\bibfnamefont {J.~C.}\ \bibnamefont {Slonczewski}},\
  }\href {\doibase 10.1103/PhysRevLett.53.190} {\bibinfo  {journal} {Physical
  Review Letters}\ ,\ \bibinfo {pages} {190}}\BibitemShut {NoStop}%
\bibitem [{\citenamefont {Chikazumi}\ and\ \citenamefont
  {Graham}(1997)}]{Chikazumi}%
  \BibitemOpen
\bibfield  {journal} {  }\bibfield  {author} {\bibinfo {author} {\bibfnamefont
  {S.}~\bibnamefont {Chikazumi}}\ and\ \bibinfo {author} {\bibfnamefont
  {C.~D.}\ \bibnamefont {Graham}},\ }\href@noop {} {\emph {\bibinfo {title}
  {{Physics of Ferromagnetism}}}},\ \bibinfo {edition} {2nd}\ ed.,\
  International Series of Monographs on Physics\ (\bibinfo  {publisher}
  {Clarendon Press},\ \bibinfo {year} {1997})\ p.\ \bibinfo {pages}
  {668}\BibitemShut {NoStop}%
\bibitem [{\citenamefont {Saitoh}\ \emph {et~al.}()\citenamefont {Saitoh},
  \citenamefont {Miyajima}, \citenamefont {Yamaoka},\ and\ \citenamefont
  {Tatara}}]{Saitoh2004}%
  \BibitemOpen
  \bibfield  {author} {\bibinfo {author} {\bibfnamefont {E.}~\bibnamefont
  {Saitoh}}, \bibinfo {author} {\bibfnamefont {H.}~\bibnamefont {Miyajima}},
  \bibinfo {author} {\bibfnamefont {T.}~\bibnamefont {Yamaoka}}, \ and\
  \bibinfo {author} {\bibfnamefont {G.}~\bibnamefont {Tatara}},\ }\href
  {\doibase 10.1038/nature03009} {\bibinfo  {journal} {Nature}\ ,\ \bibinfo
  {pages} {203}}\BibitemShut {NoStop}%
\bibitem [{Note1()}]{Note1}%
  \BibitemOpen
\bibfield  {journal} {  }\bibinfo {note} {See Supplemental Material for
  additional measurements on other samples and simulations}\BibitemShut
  {NoStop}%
\bibitem [{\citenamefont {Vansteenkiste}\ \emph {et~al.}(2014)\citenamefont
  {Vansteenkiste}, \citenamefont {Leliaert}, \citenamefont {Dvornik},
  \citenamefont {Helsen}, \citenamefont {Garcia-Sanchez},\ and\ \citenamefont
  {Van~Waeyenberge}}]{mumax3}%
  \BibitemOpen
  \bibfield  {author} {\bibinfo {author} {\bibfnamefont {A.}~\bibnamefont
  {Vansteenkiste}}, \bibinfo {author} {\bibfnamefont {J.}~\bibnamefont
  {Leliaert}}, \bibinfo {author} {\bibfnamefont {M.}~\bibnamefont {Dvornik}},
  \bibinfo {author} {\bibfnamefont {M.}~\bibnamefont {Helsen}}, \bibinfo
  {author} {\bibfnamefont {F.}~\bibnamefont {Garcia-Sanchez}}, \ and\ \bibinfo
  {author} {\bibfnamefont {B.}~\bibnamefont {Van~Waeyenberge}},\ }\href@noop {}
  {\bibfield  {journal} {\bibinfo  {journal} {AIP Advances}\ }\textbf {\bibinfo
  {volume} {4}},\ \bibinfo {pages} {107133} (\bibinfo {year}
  {2014})}\BibitemShut {NoStop}%
\bibitem [{\citenamefont {Puliafito}\ \emph {et~al.}(2014)\citenamefont
  {Puliafito}, \citenamefont {Siracusano}, \citenamefont {Azzerboni},\ and\
  \citenamefont {Finocchio}}]{puliafito2014}%
  \BibitemOpen
  \bibfield  {author} {\bibinfo {author} {\bibfnamefont {V.}~\bibnamefont
  {Puliafito}}, \bibinfo {author} {\bibfnamefont {G.}~\bibnamefont
  {Siracusano}}, \bibinfo {author} {\bibfnamefont {B.}~\bibnamefont
  {Azzerboni}}, \ and\ \bibinfo {author} {\bibfnamefont {G.}~\bibnamefont
  {Finocchio}},\ }\href@noop {} {\bibfield  {journal} {\bibinfo  {journal}
  {IEEE Mag. Lett.}\ }\textbf {\bibinfo {volume} {5}},\ \bibinfo {pages}
  {3000104} (\bibinfo {year} {2014})}\BibitemShut {NoStop}%
\end{thebibliography}%

\renewcommand{\figurename}{Sup. Fig.}
\setcounter{figure}{0} 

\begin{widetext}
\clearpage
\begin{center}
\textbf{\large{Supplemental Material}}
\end{center}
\renewcommand{\thepage}{\roman{page}}
\setcounter{page}{1}
\section*{S1 Additional measurements}

All measured samples with the same layer stack have shown almost identical results. Here we discuss measurements as a function of the angle of the applied field. We first show that the results for dc resistance and frequency resistance signal are consistent with the results presented in the main manuscript and we discuss the small differences. The nanocontact sample we show in this Supplementary Material has a nominal diameter identical to the sample from the Main manuscript but presents a slightly larger total MR of about 0.25 \% and threshold currents also slightly higher. We attribute this changes to small variations in the nanofabrication procedure.
\\
Sup. Fig.\ \ref{fig1sup} shows both the dc and high-frequency resistance response measured at the same time. We see at the onset current, a step in dc resistance and a characteristic frequency of $f\sim 20$ GHz at an applied field of $\mu_0H=$ 687 mT. As the current increases, the frequency slightly blueshifts and jumps maintaining the overall value around the 20 GHz, above the Zeeman frequency of 19 GHz
\\

\begin{figure}[htbp!]
\includegraphics[width=130 mm]{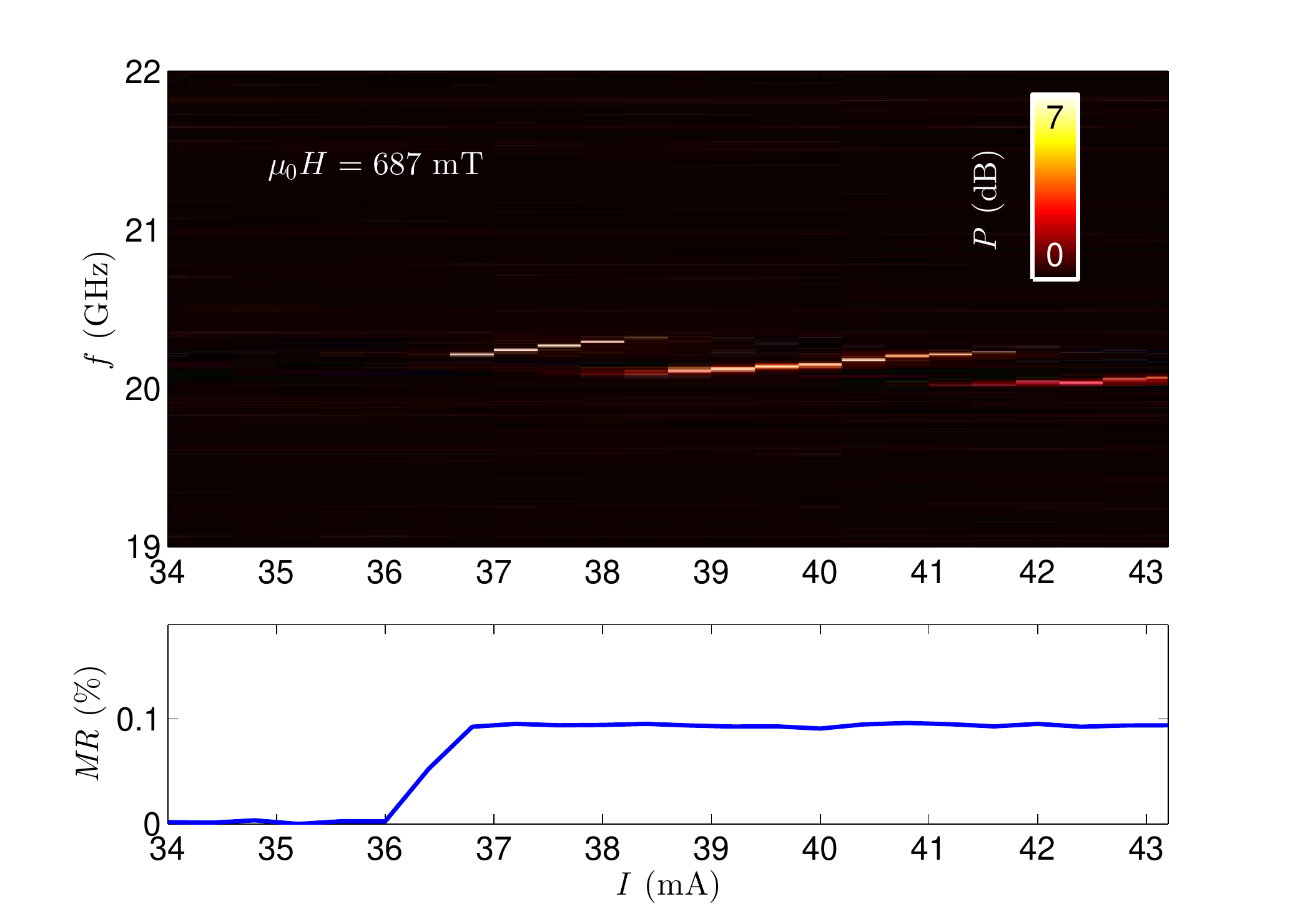}
\caption{Top panel: High-frequency spectra as a function of applied current for a field of 687~mT. Bottom panel: dc MR as a function of current for the same sample. The two measurements were taken at the same time.}
\label{fig1sup}
\end{figure}

We also measured the voltage signal at lower frequencies (hundreds of MHz). Along with the creation of droplet excitations we measure a strong and broad oscillating signal at about 300 MHz (see Sup. Fig.\ \ref{fig2sup}). All samples showing droplet solitons presented the low-frequency dynamics in the range of 100-800 MHz but the shape of the peaks were considerably different, having a well defined peak structure in some cases and a much broader structure in others.

\begin{figure}[htbp!]
\includegraphics[width=130 mm]{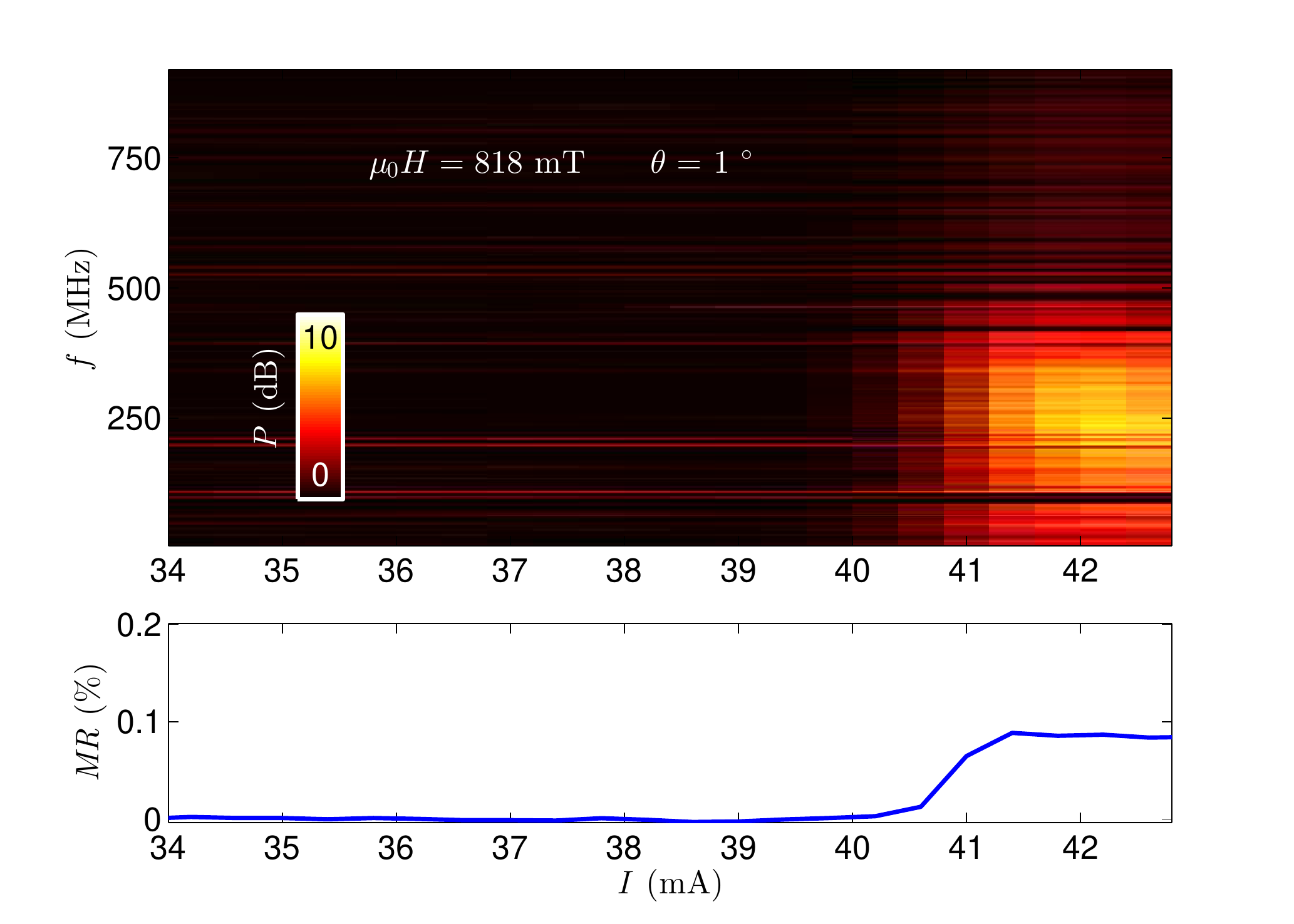}
\caption{Top panel: Low frequency spectra as a function of applied current for a field of 818~mT. Bottom panel: MR as a function of current for the same sample. The two measurements were taken at the same time.}
\label{fig2sup}
\end{figure}

Next, we measure the dependence of the droplet excitation as a function of the angle of the applied field. We first fixed the applied field strength and at each angle we measured the low-frequency spectra and the dc resistance. Droplet excitations are present only when the applied field is perpendicular to the film and we found that the maximum angle we can tilt the field before the excitation annihilates is about $15^\circ$. What we found is that as we tilt the angle the low-frequency spectra becomes stronger. Sup. Fig.\ \ref{fig3sup} shows both the dc resistance and the low frequency spectra of the nanocontact as a function of the applied field angle. We see that the low frequency spectra is larger at angles between 5 and $15^\circ$. Further, we observe in the measurements shown in Sup. Fig.\ \ref{fig3sup} the tilt in one direction produces a stronger low frequency signal. We also notice that for the stronger low-frequency signal, at positive angles between 5 and $15^\circ$ the dc MR signal also decreases a bit being consistent with the fact that the low frequency dynamics lowers the averaged resistance value.

\begin{figure}[htbp!]
\includegraphics[width=130 mm]{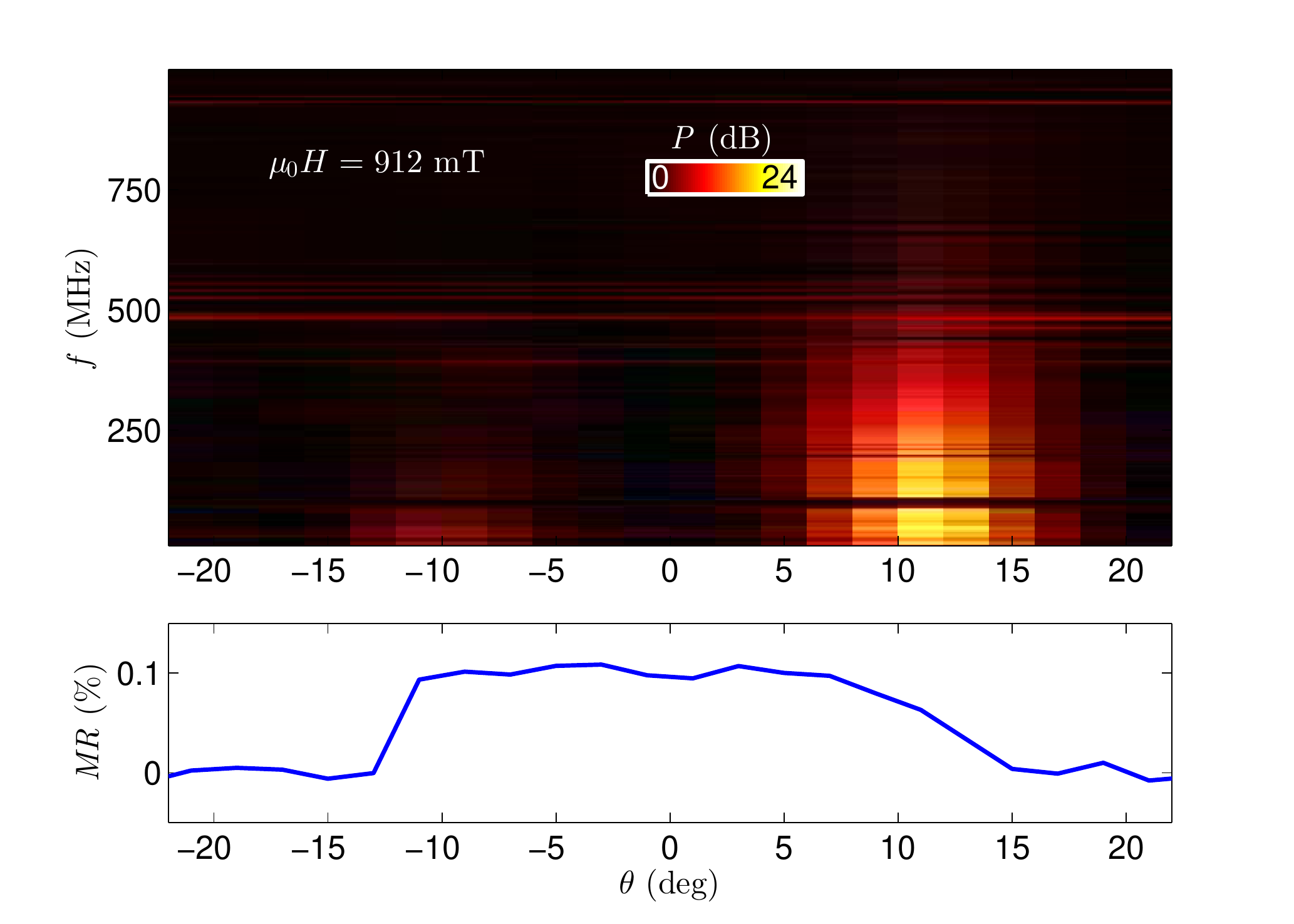}
\caption{Top panel: Low frequency spectra as a function of the angle of the applied field for a field of 912 mT. Bottom panel: MR as a function of the angle of the applied field for the same sample. The two measurements were taken at the same time for an applied current of 40 mA}
\label{fig3sup}
\end{figure}

\section*{S2 Additional micromagnetic simulations}

We present in here additional micromagnetic simulations that prove that spatial gradients in the effective field cause drift instabilities in droplet solitons. We show two representative cases, \emph{i)} a variation of 1~\% in the perpendicular applied field and \emph{ii)} a variation of about 1~\% in the anisotropy.

We have divided the nanocontact in half and considered the parameters from each part slightly different. In the first case, we applied a perpendicular field of $\mu_0H_z=1.1$ T to one half and a $\sim$ 1 \% higher in the other half ($\mu_0H_z=1.11$ T). Sup. Fig. \ref{fig4sup} shows the evolution of the droplet soliton; first, upon an applied field of 1.1 T (3 nanoseconds), and then with the slight variation of 1 \% in half of the nanocontact. We plotted a figure similar to Fig.\ 5 from the main manuscript showing how an imbalance is created in the soliton excitation (in the precession phases). We see in the upper panels of Sup. Fig.\ \ref{fig4sup} how the soliton excitation shifts in the $y$ direction until it is annihilated and a new excitation appears. The measured low frequency is very similar to the frequency we obtained in presence of an in-plane field (Fig.\ 5 shown in main manuscript).

\begin{figure}[htbp!]
\includegraphics[width=150 mm]{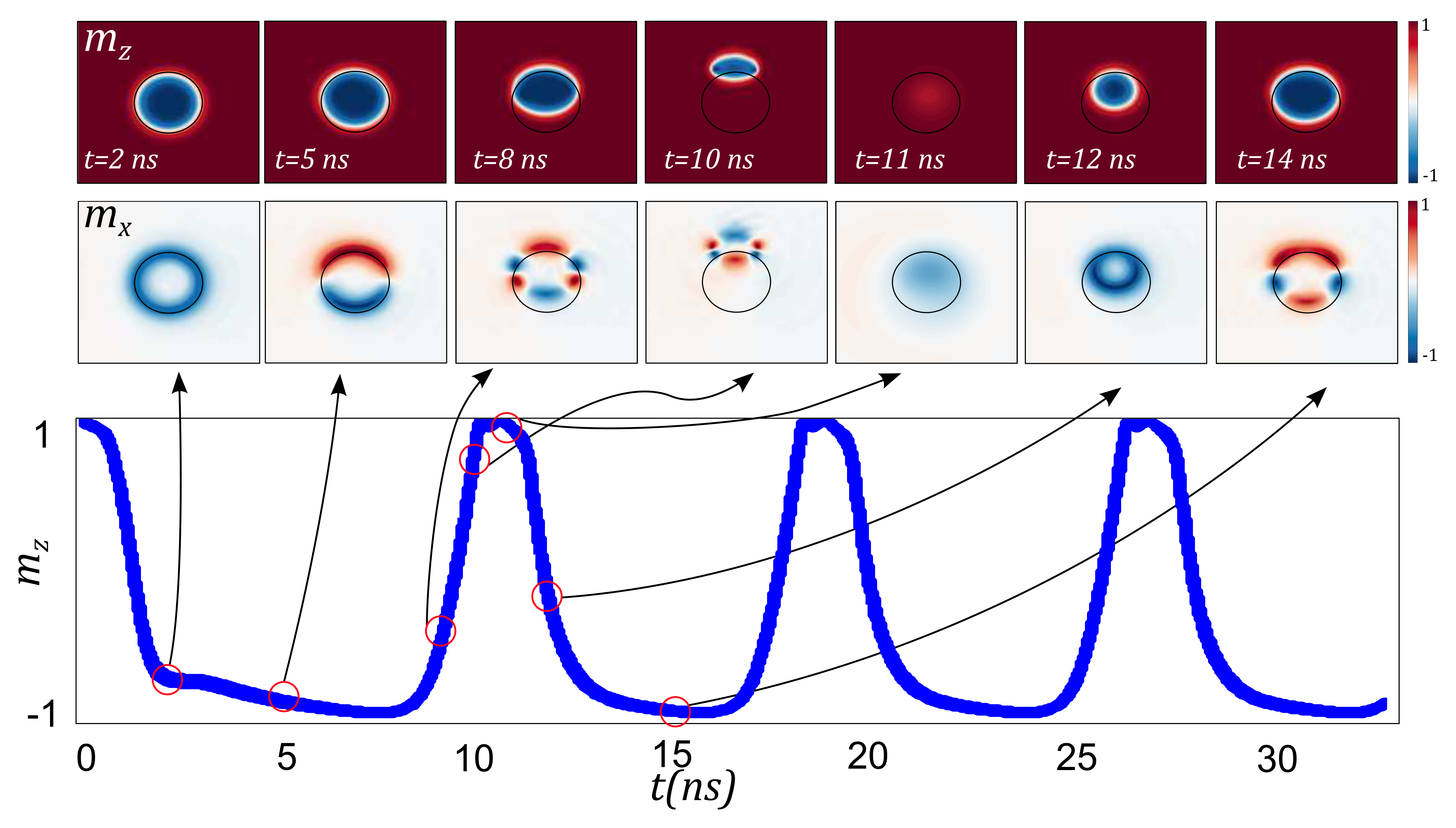}
\caption{Time evolution of the droplet soliton upon an applied field of 1.1 T perpendicular to the film plane ($t<3$ ns), and with an additional field of 0.01 T in one half of the point contact, applied at $t=3$ ns. The upper panels show magnetization maps for $m_z$, and $m_x$, at particular times of the simulation. Images correspond to a 400 $\times$ 400 nm$^2$ field view. The contact region is outlined in black. The lower panel shows the time evolution of the perpendicular component of the magnetization $m_z$, in the nanocontact area.}
\label{fig4sup}
\end{figure}

In the second case we have considered that the anisotropies of the two parts of the nanocontact differ by 1\ \%. Sup. Fig.\ \ref{fig5sup} shows the evolution of the soliton excitation upon an applied perpendicular field of $\mu_0 H_z=1.1$ T. We notice that in this case the dynamic annihilation and creation occurs at a lower frequency of about 50 MHz.
\\

\begin{figure}[htbp!]
\includegraphics[width=150 mm]{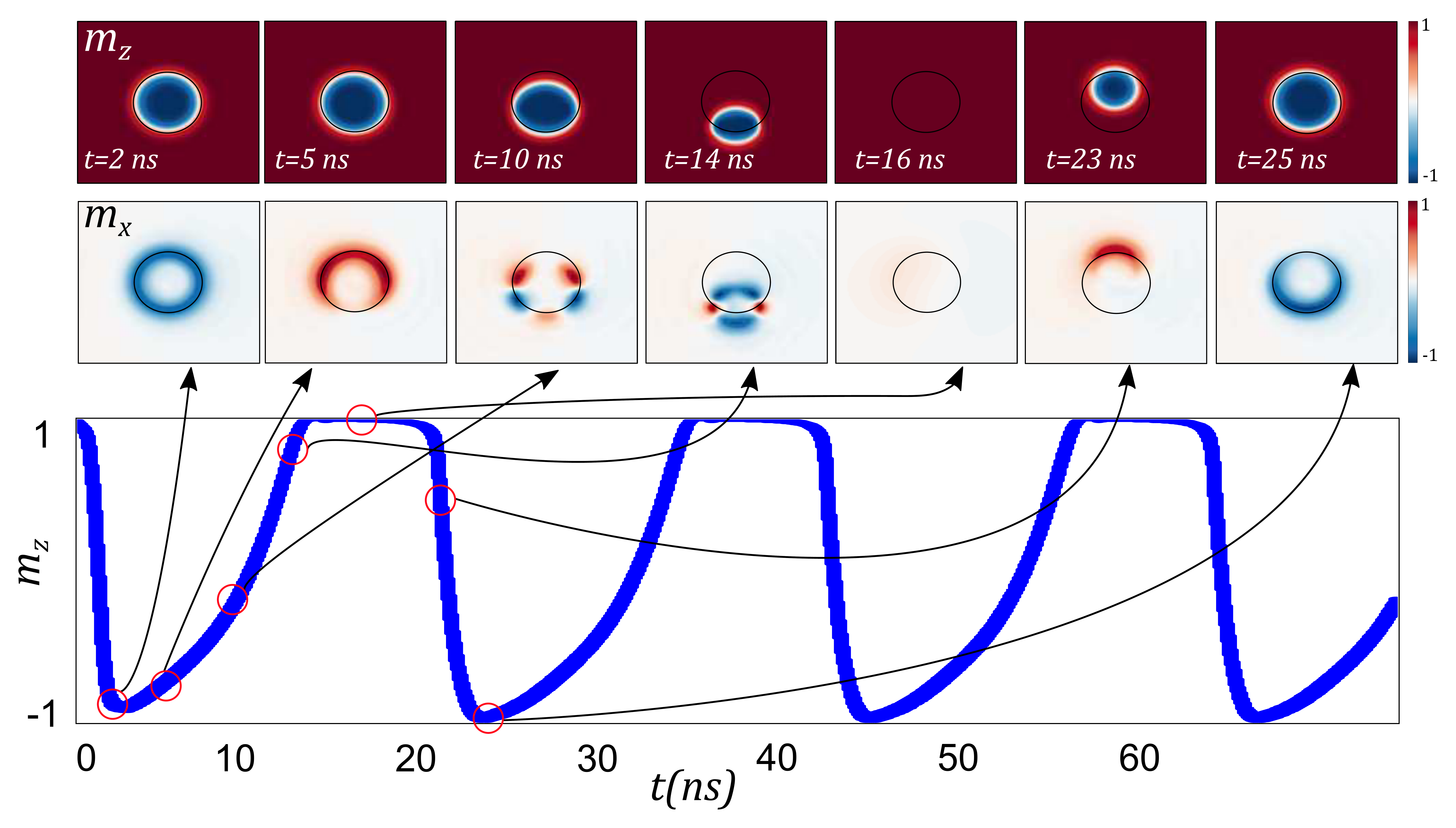}
\caption{Time evolution of the droplet soliton upon an applied field of 1.1 T perpendicular to the film plane and with an anisotropy that varies 1~\% in the two halves of the nanocontact. The upper panels show magnetization maps for $m_z$, and $m_x$, at particular times of the simulation. Images correspond to a 400 $\times$ 400 nm$^2$ field view. The contact region is outlined in blue. The lower panel shows the time evolution of the perpendicular component of the magnetization $m_z$, in the nanocontact area.}
\label{fig5sup}
\end{figure}

\newpage

\subsection*{Micromagnetics code}

\begin{spacing}{0.8}
\emph{
\footnotesize{
\noindent
//Mumax Code
\\\\
// mumax3 is a GPU-accelerated micromagnetic simulation open-source software\\
// developed at the DyNaMat group of Prof. Van Waeyenberge at Ghent University.\\
// The mumax3 code is written and maintained by Arne Vansteenkiste.\\
\\
//GRID\\
LL := 256\\
Lz := 1\\
SetGridSize(LL, LL, Lz)\\
SetCellSize(3e-9, 3e-9, 4e-9)\\
\\
//NANOCONTACT\\
diam\_circ := 150e-9\\
r\_circ := diam\_circ / 2\\
A\_circ := pi * pow(r\_circ, 2)\\
\\
DefRegion(1, layer(0).intersect(circle(diam\_circ)))\\
tableadd(m.region(1))\\
\\
//MATERIAL PROPERTIES
\\
\begin{tabular}{ll}
lambda = 1	& //Slonczewski  parameter\\
epsilonprime = 0	&// Slonczewski secondairy STT term \\
Msat = 500e3 				&//Saturation\\
Ku1 = 200e3 				&//Uniaxial Anisotropy\\
anisU = vector(0, 0, 1)&\\
Aex = 10e-12 				&//Exchange\\
alpha = 0.03 				&//Damping\\
\end{tabular}
\\
//OERSTED FIELDS--------------------\\
posX := 0.\\
posY := 0.\\
mask := newSlice(3, LL, LL, Lz)\\
current := vector(0., 0., 1.)\\
\\
for i := 0; i $<$ LL; i$++$ $\{$\\
\indent for j := 0; j $<$ LL; j$++$ $\{$\\
\indent\indent		r := index2coord(i, j, 0)\\
\indent\indent		r = r.sub(vector(posX, posY, 0))\\
\indent\indent		b := vector(0, 0, 0)\\
\indent\indent	if r.len() $>$= r\_circ $\{$\\
\indent\indent\indent			b = r.cross(current).mul(mu0 / (2 * pi * r.len() * r.len()))\\
\indent\indent		$\}$ \\
\indent         else $\{$\\
\indent\indent\indent		b = r.cross(current).mul(mu0 / (2 * pi * r\_circ * r\_circ))\\
\indent\indent		$\}$\\
\indent		for k := 0; k $<$ Lz; k$++$ $\{$\\
\indent\indent			mask.set(0, i, j, k, b.X())\\
\indent\indent			mask.set(1, i, j, k, b.Y())\\
\indent\indent			mask.set(2, i, j, k, b.Z())\\
\indent\indent		$\}$\\
\indent	$\}$\\
$\}$\\
//END OERSTED------------------------\\
\\
//INITIAL CONDITIONS\\
angle := 85.\\
my := cos(angle * pi / 180)\\
mz := sin(angle * pi / 180)\\
\\
\begin{tabular}{ll}
Value\_Bext\_z := 1.1 &//external field in z direction\\
Value\_Bext\_x := 0.15 &//external field in x direction\\
\end{tabular}\\
angle = 90.\\
px := cos(angle * pi / 180)\\
pz := sin(angle * pi / 180)\\
\\
Curr := -28e-3	\indent//current in Amps\\
Pol = 0.223\\
TableAddVar(Curr, "Current", "A")\\
\\
//SAVING\\
tableautosave(5e-13)\\
autosave(m,1e-9)\\
\\
//RUNNING\\
fixedlayer = vector(px,0.,pz)\\
m = Uniform(0, my, mz)\\
\\
\begin{tabular}{ll}
j.SetRegion(1, vector(0, 0, Curr/A\_circ*Lz))& 					//current\\
B\_ext.RemoveExtraTerms()&\\
B\_ext.add(mask, Curr) 							&				// Oersted fields\\
B\_ext = vector(0., 0, Value\_Bext\_z) 			&				//external field\\
run(3.0e-9) 									&				//running 3 ns\\ perpendicular field)\\
B\_ext = vector(Value\_Bext\_x, 0, Value\_Bext\_z)& 				// new external field\\
run(5e-8) 											&			//running 50 ns\\
\end{tabular}
}
}
\end{spacing}
\end{widetext}
\renewcommand{\thepage}{\roman{page}}
\setcounter{page}{5}

\end{document}